\documentclass[conference]{IEEEtran}
\title{Energy Model for Vesicle-based Active Transport Communication}

\usepackage{cite}

\ifCLASSINFOpdf
  \usepackage[pdftex]{graphicx}
\else
  \usepackage[dvips]{graphicx}
\fi
\usepackage[cmex10]{amsmath}
\usepackage[tight,footnotesize]{subfigure}

\usepackage{amsfonts}
\usepackage{xcolor}
\usepackage{todonotes}

\usepackage{siunitx}

\usepackage{textcomp}

\newcommand{\numAdenine}{n_{A}}
\newcommand{\numCytosine}{n_{C}}
\newcommand{\numThymine}{n_{T}}
\newcommand{\numGuanine}{n_{G}}

\newcommand{\rVesicle}{r_v}

\newcommand{\rTN}{r_{\text{TN}}}

\newcommand{\stepsize}[1]{\Delta r_{#1}}
\newcommand{\vavg}{v_{\text{avg}}}
\newcommand{\perslen}{L_p}

\newcommand{\xth}[1]{ {#1}\textrm{th} }

\newcommand{\eVesicle}{E_{S,v}}
\newcommand{\eIntraCarrying}{E_{T,\text{intra}}}
\newcommand{\eInterCarrying}{E_{T,\text{inter}}}

\begin{document}
%
\title{Energy Model for Vesicle-Based \\ Active Transport Molecular Communication}
%
%
%

\author{\IEEEauthorblockN{Nariman~Farsad\IEEEauthorrefmark{1},~\IEEEmembership{Member,~IEEE,}
        H. Birkan~Yilmaz\IEEEauthorrefmark{2},~\IEEEmembership{Member,~IEEE,}
        Chan-Byoung~Chae\IEEEauthorrefmark{2},~\IEEEmembership{Senior Member,~IEEE,}
        and~Andrea Goldsmith\IEEEauthorrefmark{1}~\IEEEmembership{Fellow,~IEEE,}}
        \IEEEauthorblockA{\IEEEauthorrefmark{1}Department of Electrical Engineering, Stanford University, CA, USA. 
        }
        \IEEEauthorblockA{\IEEEauthorrefmark{2}Yonsei Institute of Convergence Technology, School of Integrated Technology, Yonsei University, Korea. 
        }
}

%
%

\markboth{Journal of IEEE xxxx}%
{Energy Model for Vesicle-based Active Transport Communication}
%



\maketitle

\begin{abstract}
In active transport molecular communication (ATMC), information particles are actively transported from a transmitter to a receiver using special proteins. Prior work has demonstrated that ATMC can be an attractive and viable solution for on-chip applications. The energy consumption of an ATMC system  plays a central role in its design and engineering. In this work, an energy model is presented for ATMC and the model is used to provide guidelines for designing energy efficient systems. The channel capacity per unit energy is analyzed and maximized. It is shown that based on the size of the symbol set and the symbol duration, there is a vesicle size that maximizes rate per unit energy. It is also demonstrated that maximizing rate per unit energy yields very different system parameters compared to maximizing the rate only.\\
\end{abstract}


%
\IEEEpeerreviewmaketitle

\section{Introduction}

\IEEEPARstart{M}{olecular} communication (MC) is a new interdisciplinary research paradigm, where small particles such as molecules are used to convey information~\cite{Farsard_arXiv14}. At micro-scales, where the size of the transmitter and receiver are a few micrometers, the energy budgets of the nodes introduce a crucial limitation in terms of system performance\cite{kur10}. In a typical micro-scale unit, due to the device\textquotesingle s scale, energy storage can be very challenging. For example to overcome this challenge, some of the previous works have proposed energy harvesting techniques, where energy is harvested from the environment through chemical reactions \cite{freitas1999nanomedicine}. The rate at which energy can be harvested is typically a constant, and thus understanding the energy consumption of a molecular communication system plays an important role in designing the system. This work focuses on micro-scales MC in lab-on-a-chip devices, and provides an energy model that could be used for system design and optimization.   

In the literature, various MC propagation schemes have been proposed for on-chip communications, such as active transport using molecular motors \cite{dar13}, molecular communication via diffusion\cite{yilmaz2014threeDC,yilmaz2014simulationSO,yilmaz2014arrivalMF,nrkim13} and diffusion with flow \cite{far12NanoBio}, calcium signaling \cite{kuran2012calcium}, and bacterium-based communication \cite{gre10}. In \cite{far12NanoBio}, it was shown that diffusion-based molecular communication can be very slow for on-chip applications. However, active transport using stationary kinesin molecular motors and microtubule (MT) filaments, which is called active transport molecular communication (ATMC) in this work, is a viable propagation scheme for on-chip applications. In this scheme, the surface of a chip is covered with special proteins called kinesin. Kinesin mobilizes special tube like structures called microtubules (MT) through chemical reactions. This process is sometimes called kinesin-MT motility. In \cite{ekim13}, it was shown that electrical currents can be used to control the speed and direction of motion of MT filament, and in \cite{ste14} kinesin-MT motility was used for high-throughput molecular transport and assembly in microfluidics.  
In \cite{hiy09}, it was demonstrated that this technique can be used to transport vesicles in an on-chip device. The MT motility was modeled mathematically in \cite{nit06}, where it is shown that the motion of the MT in an on-chip device can be captured using Monte Carlo techniques.

Some of the prior work that have considered ATMC from a communication engineering perspective include: a complete simulation environment for ATMC in \cite{far11Bionet}, different mathematical models for the ATMC in \cite{far14TSP}, and the channel design and optimization of ATMC in \cite{far11NanoCom,far13NANO,far15TNANO}. 

In this work, an energy consumption model for ATMC system is derived. An important aspect of any communication system is the energy per bit requirements, which is the amount of energy required to transfer a single bit from the transmitter to the receiver. In \cite{kur10}, the energy model for diffusion-based MC in a 3-D environment is presented. However, to the best of our knowledge, there is no energy model proposed for ATMC systems in the MC domain. Our contributions in this work are as follows. First, a complete end-to-end energy model for ATMC systems is derived. Then using this model, system performance is analyzed by evaluating information rate per unit energy. 
Since tractable closed-form expression for the ATMC information rates do not exist, Monte Carlo simulations are used for finding the system parameter that maximize rate per unit energy. It is shown that depending on the size of the symbol set and the symbol duration, there is an optimal size for the information-carrying vesicles. The size of vesicles impact the system performance in two ways: first, they can change the probability of arrival at the receiver; second, the size of the vesicles effects the amount of energy required to create the vesicle. Finally, it is demonstrated that when maximizing rate per unit energy, the optimal system parameters may be very different from the case where only the rate is maximized. 
\begin{figure*}[t]
	\begin{center}
		\includegraphics[width=1.6\columnwidth,keepaspectratio]{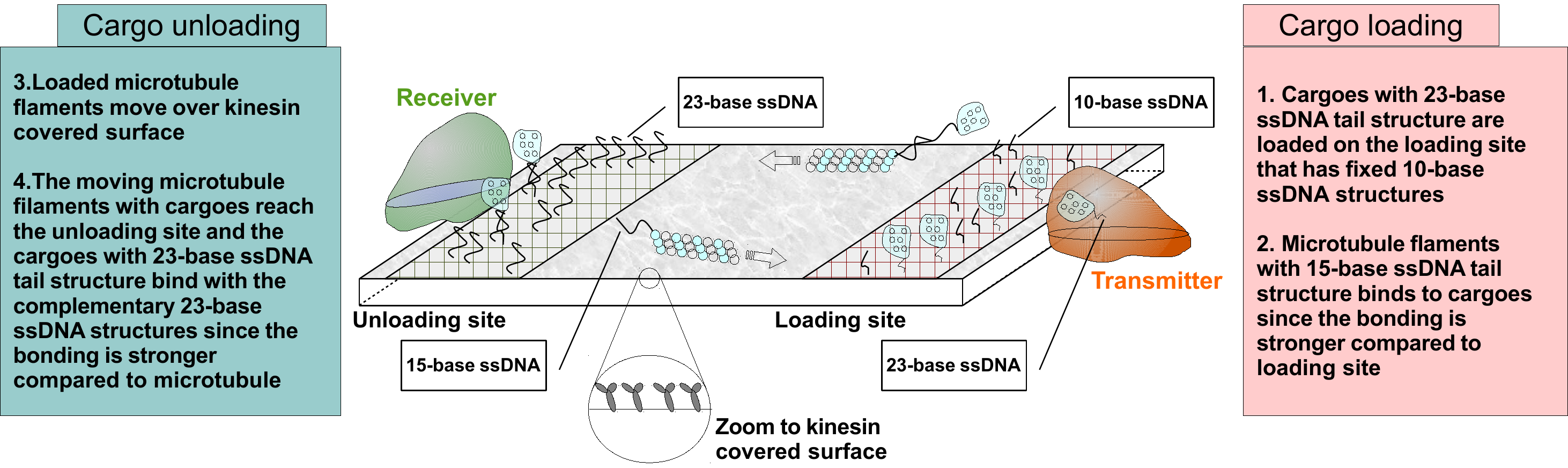}
	\end{center}
	\caption{\label{fig:MTkinDNA}Cargo transport mechanism for MT filaments gliding over stationary kinesin. Loading site contains 10-base ssDNAs, MT filaments have 15-base ssDNA tail structures, cargoes have 23-base ssDNA tail structures, and similarly unloading site contains 23-base ssDNAs. Therefore, MT filaments easily tear down the bonds between cargoes and the ssDNAs at the loading site since 15-base ssDNAs are stronger compared to 10-base ssDNAs. On the other hand, longer ssDNAs at the unloading site have the ability to trap the cargoes when the MTs with cargoes reach the unloading site.}
\end{figure*}
The rest of the paper is organized as follows. In Section \ref{sec:sysmodel}, the system model for on-chip ATMC is presented. Then, in Section \ref{sec:compEM}, energy consumption by each system component is model. An end-to-end energy model for the ATMC system is presented in Section \ref{sec:commEM}, and the problem of maximizing information rate per unit energy is formulated. Finally, in Section \ref{sec:results}, Monte Carlo simulations are used to provide design guideposts for system parameters such that information rate per unit energy is maximized.

\section{Communication System Model}
\label{sec:sysmodel}

The ATMC scheme considered in this paper is based on MT filament motility over immobilized kinesin molecular motor structures \cite{hiy09}. In nature,  MTs are one of the components of the cytoskeletal structure of the cells. They help cells to keep their structure and also act as a platform for intercellular transport. Kinesin is a molecular motor that has two ``leg-like'' structures, which are used to ``walk'' over MT tracks. This process is powered by adenosine triphosphate (ATP) hydrolysis reaction, where ATP is converted into adenosine diphosphate (ADP). Inside cells MTs act as tracks and kinesin as locomotives transporting cargoes.

In \cite{how89}, it was shown that a single stationary kinesin can move an MT filament. Note that the moving entity is the MT filament not the molecular motor structure. Later it was shown that kinesin covered substrates can create MT motility in an on-chip device, and can also be used to transport cargoes from one location to another \cite{hiy09}. In \cite{far11NanoCom}, it was demonstrated that using information particles as cargoes, a complete MC system can be implemented based on ATMC.

In Fig.~\ref{fig:MTkinDNA}, the main entities of an ATMC system are depicted: the transmitter node, glass substrate that is covered with stationary kinesin, cargo vesicles that have single-stranded DNA (ssDNA) tails, MT filaments, and the receiver node. The transmitter node, which is assumed to be a synthetic cell, is responsible for generating the vesicles. These vesicles are roughly spherical and have a similar structure as the cell membrane. The walls of the vesicles are generally constructed from phospolipids. The glass substrate, from a communication perspective, has two special regions: loading and unloading zones. The transport mechanism between these zones uses DNA hybridization and stationary kinesin that enables MT motility.

\subsection{DNA Hybridization \& Transport Mechanism of ATMC}
Generally, DNA molecules are composed of two polymer strands that bond together in a helical fashion through non-covalent bonds. The two strands can come apart to form two ssDNA molecules. In molecular biology, hybridization is a phenomenon in which ssDNA molecules anneal to DNA. Hence, complementary ssDNA structures can be used for selective hybridization to engineer the vesicle anchoring, pick up and drop off processes. Table~\ref{tab_ssdna_structures} shows the ssDNA sequences attached to the different components of the system.

\begin{table}[th]
\caption{The ssDNA sequences used in different parts of the ATMC~~\cite{hiy09}}
\label{tab_ssdna_structures}
\renewcommand{\arraystretch}{1.2}
\centering
\begin{tabular}{ll}
\hline  
 Location &  ssDNA sticky end region sequence \\
\hline 
Cargo Tail     & \textbf{TGT} \textbf{GCAACACTAC} \textbf{AATCAGCGAA} \\
Loading Site    & \textbf{TTCGCTGATT} \\
MT (cargo carrier) & \textbf{TTCGCTGATT} \textbf{GTAGT} \\
Unloading Site  &  \textbf{TTCGCTGATT} \textbf{GTAGTGTTGC} \textbf{ACA}\\
\hline
\end{tabular}
\end{table}

Note that the length of the ssDNAs attached to the loading site is designed to be much shorter than the ones attached to the cargo. When a cargo is anchored to an ssDNA at the loading site, there are 13-bases that are not connected. This partial DNA hybridization would allow the MT to load the vesicle from its anchoring position at the loading site when it moves close to the vesicle. The main steps of the transport mechanism are listed below: 
\begin{enumerate}
\item vesicles/cargoes with 23-base ssDNA tails are anchored to the loading site that is covered with 10-base ssDNA structures
\item MT filaments covered with 15-base ssDNA pick up the vesicles when they glide close to it.
\item vesicles/cargoes are unloaded at the unloading site, which is covered with 23-base ssDNA structures, when a loaded MT filament glides by.
\end{enumerate}

\subsection{Propagation Dynamics}
Although this form of active transport has been implemented and tested in wet labs \cite{hiy09}, it is very difficult to obtain the statistical data that are needed for studying the communication engineering aspects of the system using experimentation. Therefore, previous works have relied on computer simulations to study these systems~\cite{far14TSP,far11NanoCom,far12NanoBio}. In this work, the simulation environment presented in~\cite{far11NanoCom} is utilized.

In the simulation environment, the MT filaments move only in the x-y directions and do not move in the z direction (i.e., they move on the planar glass substrate and do not move along the height of the channel). This is a realistic representation of the motion of MT, since they always move right on top of kinesin-covered surface. During each time interval $\Delta t$, the MT filaments travel to a new coordinates according to
\begin{align}
\begin{split}
x_{i} &= x_{i\!-\!1} + \stepsize{i} \, \cos \theta_{i} \\
y_{i} &= y_{i\!-\!1} + \stepsize{i} \, \cos \theta_{i} 
\end{split}
\end{align}
where $x_i$, $y_i$, $\stepsize{i}$, and $\theta_{i}$ correspond to the x and y coordinates, step size, and angle of direction of motion at the $\xth{i}$ time interval, respectively. At each step, a sample of $\stepsize{i}$ is generated acoording to a Gaussian distribution with the following parameters
\begin{align}
\stepsize{i} &\sim \mathcal{N}(\vavg \Delta t, \,\, 2D\Delta t)
\end{align}
where $\mathcal{N}(\mu, V)$, $\vavg$, and $D$ correspond to the Gaussian distribution with mean $\mu$ and the variance $V$, the average velocity of the MT, and the diffusion coefficient of the MT, respectively. The angle $\theta_i$ is the sum of the angle in the previous step and the angular change, where the angular change $\Delta \theta_i$ is also a Gaussian random variable given as
\begin{align}
\begin{split}
\theta_i &= \theta_{i\!-\!1} + \Delta \theta_i \\
\Delta \theta_i &\sim \mathcal{N}(0, \,\, \vavg\Delta t/\perslen)
\end{split}
\end{align}
where $\perslen$ corresponds to the persistence length of the MT's trajectory. These system parameters are typically ${\vavg = \SI{0.85}{\micro\meter/\second}}$ and ${\perslen = \SI{111}{\micro\meter}}$ as was shown in~\cite{nit06}. 

The MT filaments are assumed to be without a cargo and placed randomly on the glass substrate. Moreover, the initial direction of motion for each MT is selected randomly from the range $[0,2\pi]$. From experimental observations, an MT loads a vesicle only if it passes in close proximity of the vesicle in the loading zone~\cite{hiy09}. Therefore, the loading zone is divided into squares forming a grid, where each square in the grid has the same length as the diameter of the vesicles. In the simulations, it is assumed that an MT loads a vesicle if it passes the square containing the vesicle. The reader is referred to~\cite{far11NanoCom} for a more detailed explanation of the simulation environment.

\section{Energy Model for ATMC Components}
\label{sec:compEM}

In this paper, we consider the MC systems in which the communication nodes are able to produce and store energy. Some of the produced energy is used by routine activities of the unit, while the rest is available for the communication processes. In the rest of this work, only the energy consumed by the communication system is considered. 

As shown in Fig.~\ref{fig:MTkinDNA}, it is assumed that the transmitter and the receiver modules are synthetic cells. In our communication model, vesicles are used to carry information, and it is assumed that information is encoded in the number of vesicles released by the transmitter.

In a vesicle-based active transport system, 
energy is spent for the following steps: vesicle synthesis by the synthetic transmitter cell; carrying vesicles from inside the synthetic transmitter cell to the loading zone; anchoring the vesicles at the loading zone; loading vesicles to MTs; MT motility and vesicle transport; unloading the vesicles at the unloading zone.


The first step in the communication process is vesicle synthesis at the synthetic transmitter cell. As mentioned before, vesicles are hollow spheres whose walls are constructed out of a variety of structures, most notably phospholipids. The cost of synthesizing one phospholipid molecule is 1 unit of ATP \cite{lehninger2000principles}, which in turn equals $\SI{83}{\zepto\joule}$~\cite{freitas1999nanomedicine}. In a secretory vesicle, there are 5 phospholipids in $\SI{1}{\nano\meter^2}$~area \cite{freitas1999nanomedicine}. Thus, the total cost of synthesizing a vesicle, $\eVesicle$, with a radius of $\rVesicle$ is
\begin{equation}
\eVesicle = \frac{5}{\si{\nano\meter^2}} \, (4\pi\rVesicle^2) \,\, \SI{83}{\zepto\joule}.
\end{equation}


Next step is for the transmitter to carry the vesicle from the internal synthesis location to the loading zone. The vesicle moves along MT tracks to the membrane of the synthetic transmitter cell with the help of the motor protein kinesin. As was explained in the previous section, ATP hydrolysis reaction creates the kinetic energy and causes the kinesin to ``walk'' on MT tracks.  For each ATP molecule used by kinesin, it travels $\SI{8}{\nano\meter}$ carrying the vesicle with it~\cite{alberts2007molecular}. It is assumed that the total distance that needs to be traveled is roughly half of the transmitter node's radius. Hence, the overall cost of this intranode transportation, $\eIntraCarrying$, per vesicle is
\begin{equation}
\eIntraCarrying = \left\lceil \frac{\rTN/2}{\SI{8}{\nano\meter}} \right\rceil \SI{83}{\zepto\joule}
\end{equation}
where $\rTN$ denotes the radius of the transmitter node.

The vesicles released by the transmitter will be trapped at the loading zone via DNA hybridization. Note that the length of an ssDNA attached to the loading zone is much shorter than that attached to the cargo. Therefore, MTs with longer ssDNA strands can easily pick up the cargoes. Energy cost of hybridization depends on the length of the ssDNA and the bond energies associated with different bases Adenine, Thymine, Guanine, and Cytosine. Adenine and Thymine have two hydrogen bonds while Guanine and Cytosine have three hydrogen bonds, where the energy of each hydrogen bond is \SI{35}{\zepto\joule} \cite{freitas1999nanomedicine}. Thus, the energy of these bonds is given by
\begin{equation}
E_H = 35 \, [2 (\numAdenine+\numThymine) + 3 (\numCytosine + \numGuanine)] \, \si{\zepto\joule},
\end{equation}
where $\numAdenine$, $\numThymine$, $\numGuanine$ and $\numCytosine$ represent the number of Adenine, Thymine, Guanine, and Cytosine in the ssDNA, respectively. For the system considered in this work, the ssDNA sequences for each component is given in Table~\ref{tab_ssdna_structures}. The length of the used ssDNAs is 10 base-pairs ($\numAdenine \!+\! \numThymine \!=\! 6$, $\numCytosine \!+\! \numGuanine \!=\!4$) for anchoring the information particle at the loading zone , 15 base-pairs ($\numAdenine \!+\! \numThymine \!=\! 9$, $\numCytosine \!+\! \numGuanine \!=\!6$) for attaching to the MT and 23 base-pairs ($\numAdenine \!+\! \numThymine \!=\! 13$, $\numCytosine \!+\! \numGuanine \!=\!10$) for drop off at the unloading zone. Therefore, the energy required for anchoring, loading and drop-off are
\begin{align}
E_{HA} &= 35 \, [2 (6) + 3 (4)] \, \si{\zepto\joule}= 840\, \si{\zepto\joule}, \\
E_{HL} &= 35 \, [2 (9) + 3 (6)] \, \si{\zepto\joule}= 1260\, \si{\zepto\joule}, \\
E_{HD} &= 35 \, [2 (13) + 3 (10)] \, \si{\zepto\joule}= 1960\, \si{\zepto\joule}, 
\end{align}
respectively.

Next, the MT moves on the kinesin covered surface and the loaded cargoes are transported by motile MTs gliding over kinesin covered substrate to the unloading site. As was mentioned earlier, the motor protein kinesin takes a single step along the MT for each ATP molecule it hybridizes, where the distance traveled is about $\SI{8}{\nano\meter}$ on average~\cite{coy99}. To calculate the number of ATP molecules used by kinesin per unit time, the average speed of kinesin is used. In \cite{coy99}, this value was observed to be $\SI{800}{\nano\meter/\second}$. Therefore, on average the kinesin goes through 100 ATP hydrolysis reactions per second. Let $\sigma_k$ be the density of kinesin molecules on the substrate. Typically, this density can be between 50 per $\si{\micro\meter^2}$ to 100 per $\si{\micro\meter^2}$ \cite{gib01}. Let $L_m$ be the average length of MTs. Then, the overall cost of MT motion for time duration $t$, $\eInterCarrying (t)$, is
\begin{equation}
\eInterCarrying (t) =  100 t \, L_m\sqrt{\sigma_k}\, \SI{83}{\zepto\joule}
\end{equation}
where $L_m\sqrt{\sigma_k}$ is the average number of kinesin that are in contact with the MT filament at any given time. Note that this is a good estimate for this value since the MT is $\SI{24}{\nano\meter}$ in diameter and can be regarded as a line with no width.

\section{Energy Model for Communication}
\label{sec:commEM}
An end-to-end energy model for the ATMC channel is presented in this section. It is assumed that the information is conveyed through the number of vesicles transported. Even if the information-carrying molecules are inside the vesicles, the more vesicles transported, the more information transported. Therefore, this assumption is valid for this channel.

Let $X$ be the number of vesicles generated and released by the transmitter, and let $Y_\tau$ be the number that is received after a symbol interval $\tau$. Let the probability mass function (PMF) $P(x)$ be the input distribution, and the conditional PMF $P(y_\tau|x)$ represent the channel. The average total energy used by this communication system is 
\begin{align}
\begin{split}
E_{ATMC} &= (\eVesicle+\eIntraCarrying+E_{HA}
)\mathbb{E}[X] \\
         &\;\;+\eInterCarrying (\tau)M+(E_{HL}+E_{HD})\mathbb{E}[Y_\tau],
\end{split}
\end{align}
where $\eVesicle$, $\eIntraCarrying$, $E_{HA}$, $\eInterCarrying$, $E_{HL}$, and $E_{HD}$ are the energy models for vesicle generation, intracellular transport, transmission zone attachment, MT motion, vesicle loading, and unloading given in the previous section. The $\mathbb{E}[.]$ is expectation, where both expectations can be calculated from $P(x)$ and $P(y_\tau|x)$, and $M$ is the number of MTs in the channel.  

From the expression for the total energy, it is observed that three system parameters have a significant effect on the energy consumption of the ATMC system: the size of the vesicles $r_v$, the symbol interval (i.e. time per channel use) $\tau$, and the input symbol set size $|\mathcal{X}|$ and its PMF $P(x)$. Our goal in this work is to provide design guidelines for an ATMC system with limited energy budget. This problem is formulated as
\begin{align}
	\max_{P(x),|\mathcal{X}|,\tau,r_v} \frac{I(X;Y_\tau)}{E_{ATMC}},
\end{align}
where $I(X;Y_\tau)$ is mutual information between the channel input and the output. Since there are no closed-form expressions for mutual information for ATMC channel \cite{far14TSP}, Monte Carlo simulations are used to provide design guidelines for the system parameters that maximize this term. In particular, in \cite{far12NanoBio}, it is shown that Monte Carlo simulations could be used to estimate $P(y_\tau|x)$. The Blahuat-Arimoto \cite{bla1972,ari72} algorithm is then employed to find the input probability distribution $P(x)$ that maximizes mutual information. Then, different system parameters are changed and the effects on information rate per unit energy are observed. In the next section, the simulation results are presented, and optimal design strategies discussed. In the simulations, it is assumed that the channel is memoryless. This assumption can be justified in two ways. First, the symbol intervals can be made large enough to reduce and remove inter-symbol interference. Second, it is possible to use chemical reactions (e.g. using enzymes) to destroy the vesicles that remain in the channel \cite{kuran2013tunnelBA,noe14,heren15}. Relaxing this assumption and studying the resulting system is future work. 

\section{Results \& Energy Limited Design}
\label{sec:results}
\begin{figure}[t]
	\begin{center}
		\includegraphics[width=0.95\columnwidth,keepaspectratio]{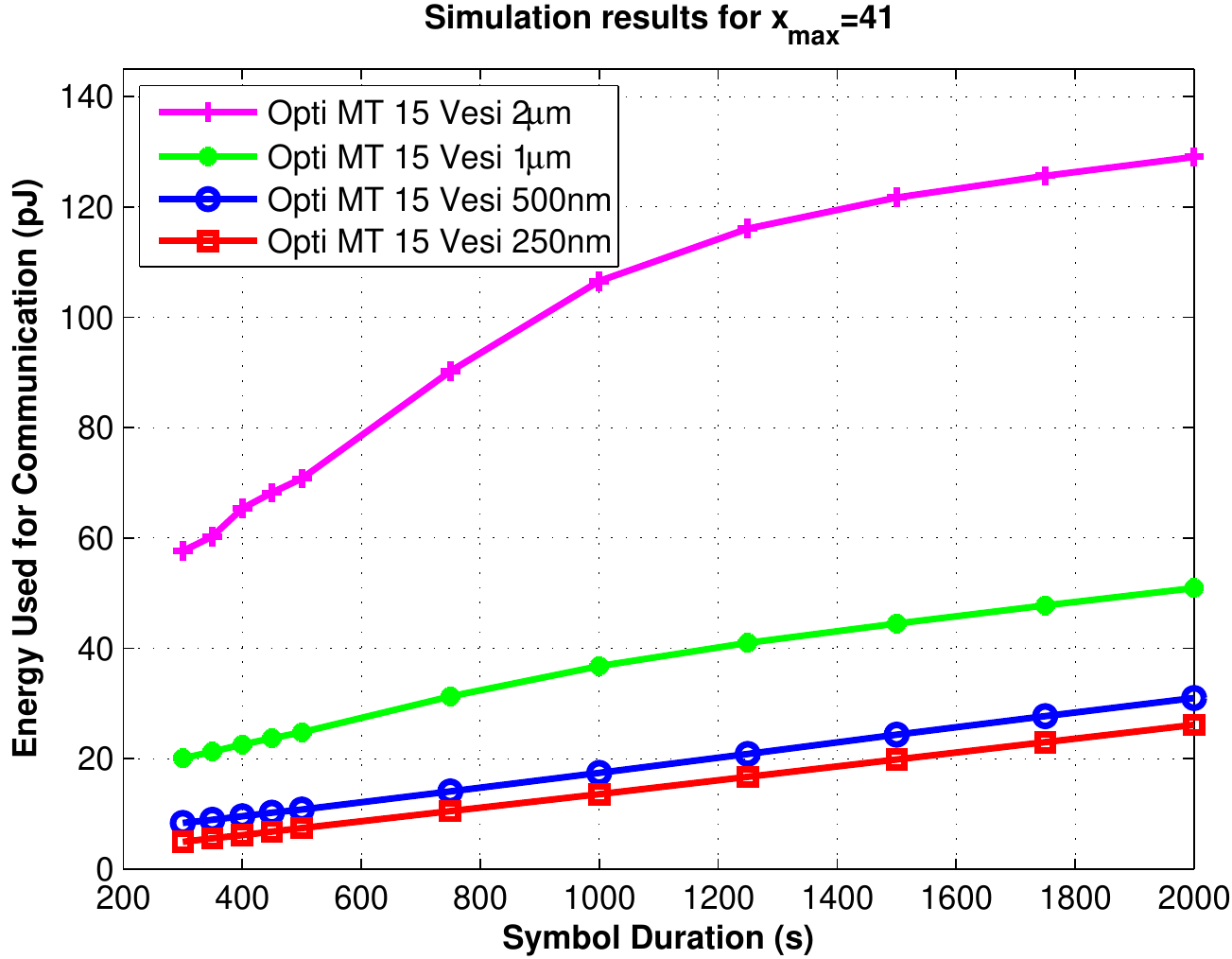}
	\end{center}
	\caption{\label{fig:EnrgVsd} {The energy used during one symbol duration by the communication system in pico-joules versus the symbol duration. The cardinality of the symbol set is $x_{max}=41$. The energy consumption increases as the symbol duration increases.}}
\end{figure}

Monte Carlo simulation are used in this section to study the effects of different system parameters on the total energy expenditure, and to provide guideposts for energy efficient design that maximizes information rate per unit energy. In the simulations, it is assumed that the input symbol set is $\mathcal{X}$, where the elements are the number of vesicles generated and released by the transmitter. Let $x_{\max}=|\mathcal{X}|$ be the cardinality of the message set. Then at the beginning of each symbol duration (i.e. time per channel use), $X\in\mathcal{X}$ vesicles are released by the transmitter. In the simulations, the channel dimensions are assumed to be $\SI{20}{\micro\meter} \times \SI{60}{\micro\meter}$, where the separation distance between the transmitter and the receiver is $\SI{40}{\micro\meter}$. It is also assumed that the transmitter is placed in its optimal location according to \cite{far11NanoCom}, the length of the MT is $\SI{10}{\micro\meter}$, there are 15 MTs in the channel, and each MT can load multiple vesicles, up to a maximum. This maximum is assumed to be equal to the half the length of the MT divided by the diameter of the vesicles. All these parameters are based on laboratory experimentation in \cite{hiy09}.

First, we analyze the effects of the symbol duration and the size of the vesicles. Fig.~\ref{fig:EnrgVsd} shows the energy used by the communication system as a function of symbol duration. The four curves show the results for different sized vesicles with diameters of 250 nm, 500 nm, $\SI{1}{\micro\meter}$, and $\SI{2}{\micro\meter}$. The size of the symbol set is fixed at 41 (i.e. the number of vesicles that could be released during each symbol duration is between 0 to 40). 
From the figure it is observed that the energy increases linearly with symbol duration and non-linearly with the size of the vesicle. The energy consumption tends to be in the order of a few pico-joules to a few hundred pico-joules. 

\begin{figure}
	\begin{center}
		\includegraphics[width=0.95\columnwidth,keepaspectratio]{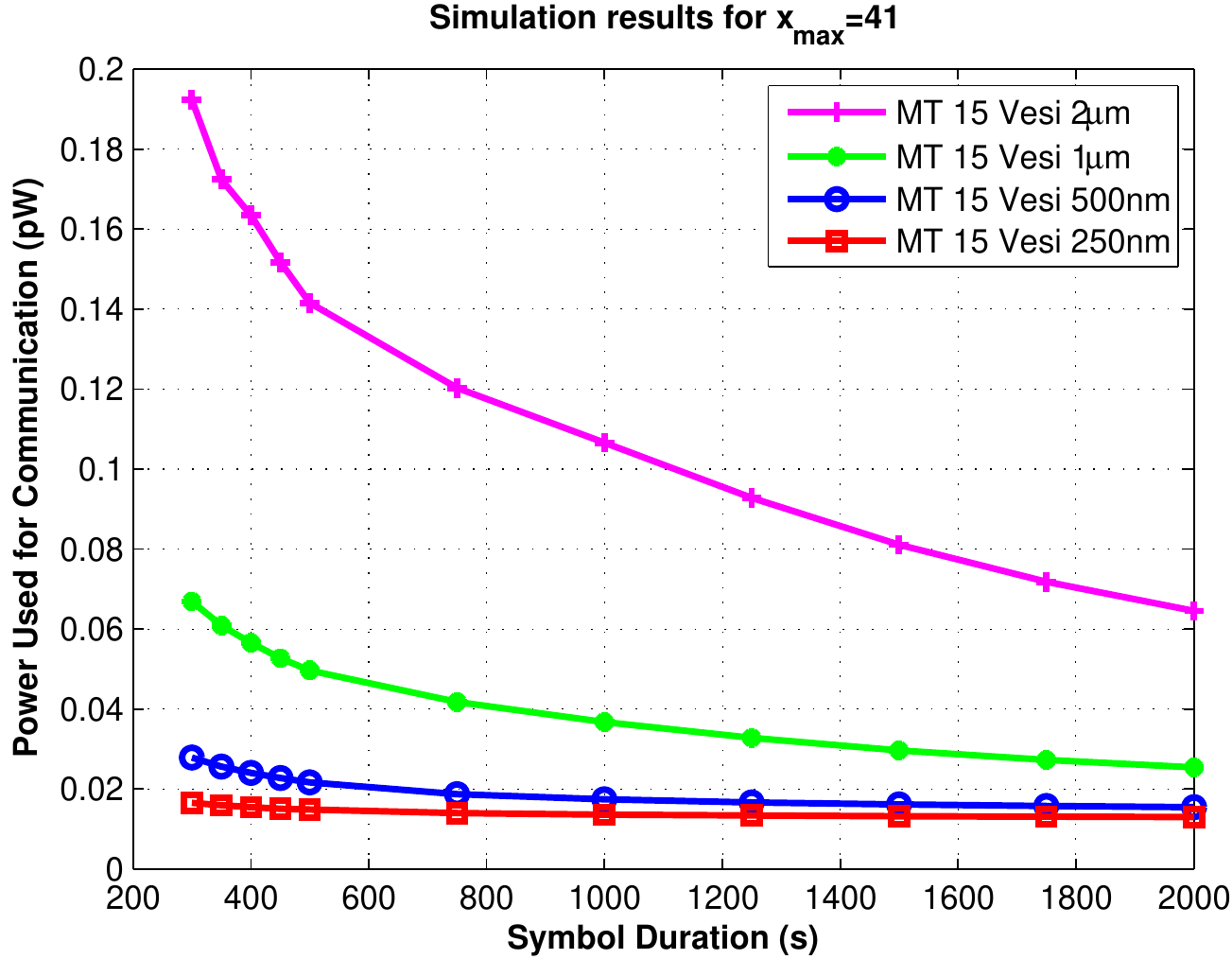}
	\end{center}
	\caption{\label{fig:pwrVsd} {The power used by the communication system in pico-watts versus the symbol duration. The cardinality of the symbol set is $x_{max}=41$s. Interestingly, the power decreases as the symbol duration increases.}}
\end{figure}

Figure~\ref{fig:pwrVsd} shows the power used by the communication system in pico-watts versus the symbol duration. Interestingly, as the symbol duration increases, the power used by the system decreases. This is due to the fact that as the symbol duration increases, the interval between vesicle generation between consecutive symbols increases, and hence the power decreases. The power used by the communication system is on the order of a few tens of femto-watts to less than half a pico-watt.

\begin{figure}
	\begin{center}
		\includegraphics[width=0.95\columnwidth,keepaspectratio]{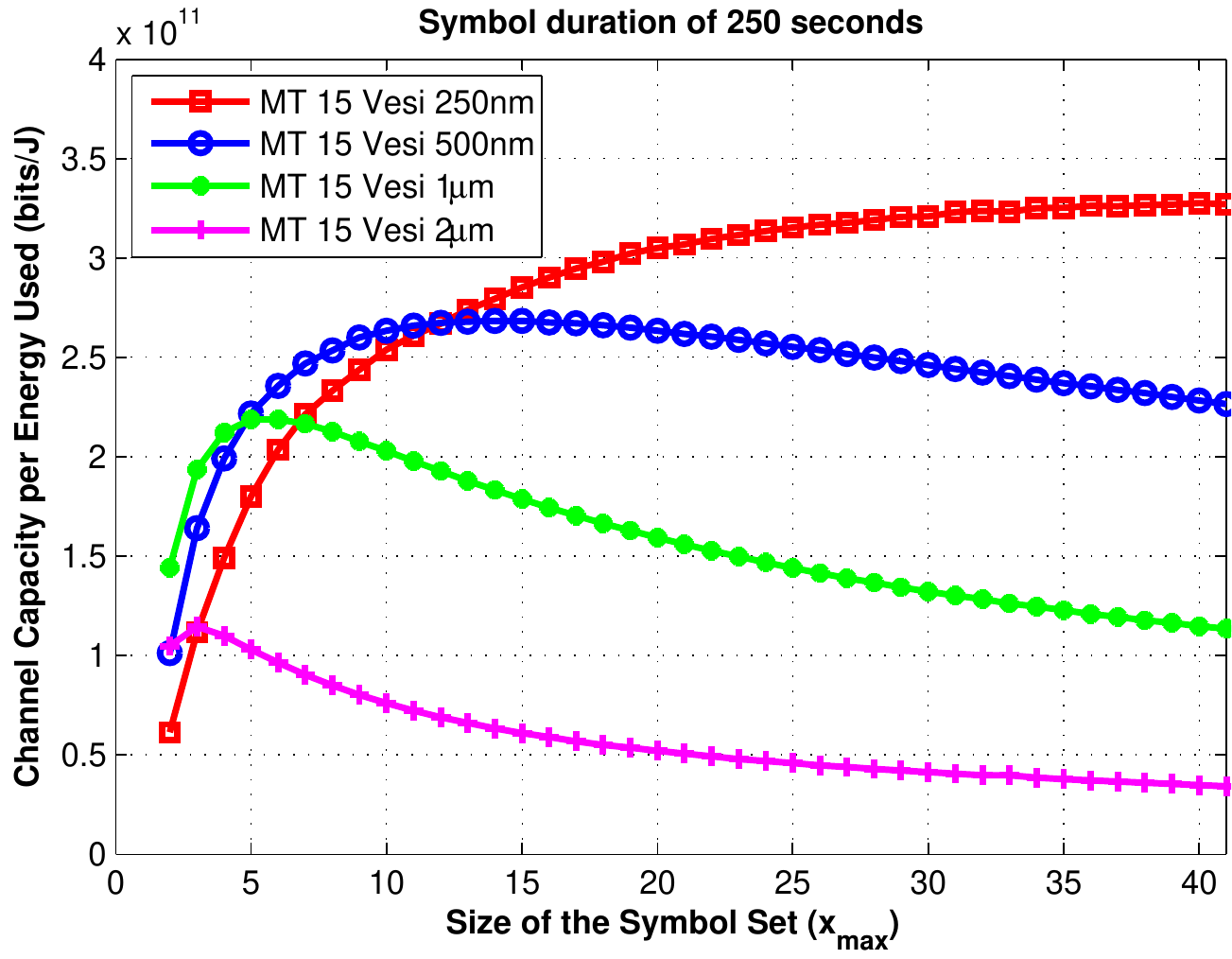}
	\end{center}
	\caption{\label{fig:capPerEngVsss} {The channel capacity per energy used in bits per joule versus the size of the symbol set. The symbol duration is 250 seconds. When energy efficiency is considered, there is an optimal symbol set size.}}
\end{figure}

\begin{figure}
	\begin{center}
		\includegraphics[width=0.97\columnwidth,keepaspectratio]{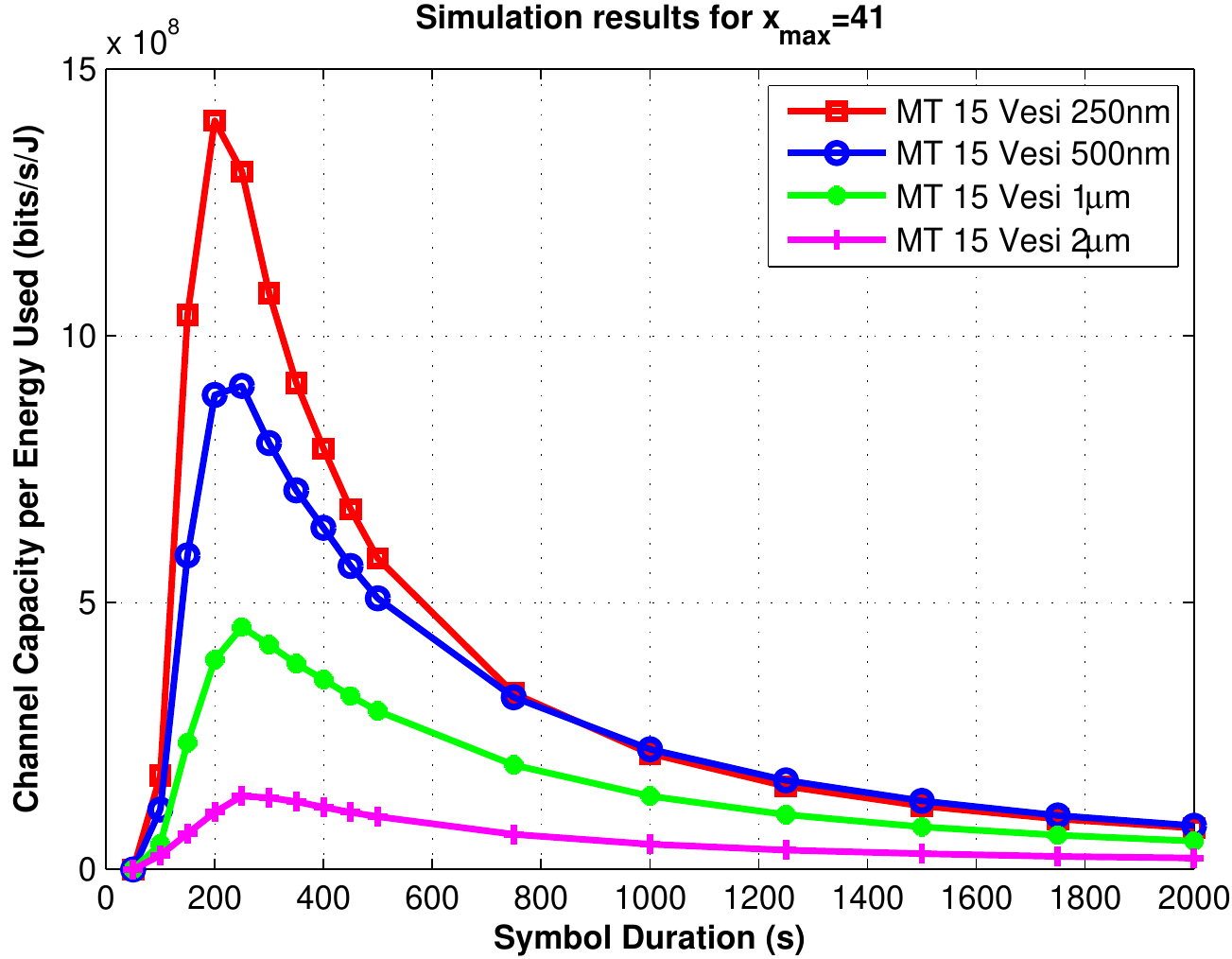}
	\end{center}
	\caption{\label{fig:capPersecPerEngVsd} {The channel capacity per energy used in bits per second per joule versus the symbol duration. The symbol set size is $x_{max}=41$. When energy efficiency is considered, the use of smaller vesicles tend to be optimal.}}
\end{figure}

From the total energy expression derived in the previous section, as the size of the symbol set increases, it is expected that the total energy consumption increases. Fig.~\ref{fig:capPerEngVsss} shows the channel capacity per energy used in bits per joules as a function of the size of the symbol set. It is observed that unlike previous works that showed larger symbol sets are desirable \cite{far12NanoBio}, when optimizing the energy there is an optimal symbol set size. Moreover, depending on the size of the symbol set and the symbol duration, there is an optimal vesicle size that maximizes the capacity per unit energy.

Finally, in Fig.~\ref{fig:capPersecPerEngVsd} the channel capacity per unit energy is presented in units of bits per second per joule as a function of symbol duration. The size of the symbol set is constant at $x_{max}=41$. In \cite{far11NanoCom}, it was shown that the $\SI{1}{\micro\meter}$ vesicle size maximizes information rate, while 250 nm vesicle size achieves the lowest rate compared to the other vesicle sizes. However, when capacity per unit energy is considered, the 250 nm vesicles achieves the highest capacity for these particular system parameter. This shows that when designing molecular communication channels, energy is an important parameter that must be considered in the design process.

\section{Conclusion}
An energy model for active transport molecular communication (ATMC) was presented in this work. It was shown that the energy consumed by the communication system depends on the size of the vesicles, the symbol duration, and the input symbol set size and its probability mass function. To evaluate the effects of these parameters on energy consumption and information rate per unit energy, Monte Carlo simulations were employed. From the results, it was shown that for a given symbol set size and symbol duration, there is a vesicle size that maximizes rate per energy, and similarly, for a given vesicle size and symbol duration there is a symbol set size that maximizes rate per unit energy. An interesting property that was observed is the significant difference between maximizing rate and maximizing rate per unit energy. As was shown in \cite{far11NanoCom}, when only the rate is maximized, larger vesicle sizes and larger symbol set sizes are desirable. However, when rate per unit energy is maximized, smaller vesicle sizes and smaller message sets are preferred. Therefore, when designing molecular communication systems energy can have a significant effect on the overall design.

\section*{Acknowledgment}
This research was supported in part by the NSF Center for Science of Information (CSoI) under grant CCF-0939370,  the NSERC Postdoctoral Fellowship fund PDF-471342-2015, the MSIP (Ministry of Science, ICT and Future Planning), Korea, under the ``IT Consilience Creative Program" (IITP- 2015-R0346-15-1008) supervised by the IITP (Institute for Information \& Communications Technology Promotion) and by the Basic Science Research Program (2014R1A1A1002186) funded by the Ministry of Science, ICT and Future Planning (MSIP), Korea, through the National Research Foundation of Korea.


\bibliographystyle{IEEEtran}
\bibliography{IEEEabrv,energyModelATMC}
%

%








\end{document}